\documentclass[12pt,preprint]{aastex}

\shorttitle{Triple Main Sequence in NGC~2808}
\shortauthors{Piotto et al.}

\begin{document}
  \title{A Triple Main Sequence in the Globular
          Cluster NGC 2808\thanks{Based on observations
          with the NASA/ESA {\it Hubble Space Telescope},
          obtained at the Space Telescope Science Institute, which is
operated
          by AURA, Inc., under NASA contract NAS 5-26555. } }

\author{G.\ Piotto\altaffilmark{1},
        L.\ R.\ Bedin\altaffilmark{2},
        J.\ Anderson\altaffilmark{3},
        I.\ R.\ King \altaffilmark{4},
        S.\ Cassisi\altaffilmark{5},
        A.\ P.\ Milone\altaffilmark{1},
        S.\ Villanova\altaffilmark{1},
        A.\ Pietrinferni\altaffilmark{5},
        A.\ Renzini\altaffilmark{6}
        }

% KK: shortened format of these
%\altaffiltext{1}{Dipartimento di Astronomia, Universit\`a di Padova,
%                 Vicolo dell'Osservatorio 3, I-35122 Padua, Italy\\
%                \email{giampaolo.piotto@unipd.it,sandro.villanova@unipd.it}}
%\altaffiltext{2}{European Southern Observatory, Karl-Schwarzschild-Strasse 2,
%                 D-85748 Garching, Germany\\
%                 \email{lbedin@eso.org}}
%\altaffiltext{3}{Department of Physics and Astronomy, Mail Stop 108, Rice
%                 University, 6100 Main Street, Houston, TX 77005\\
%                 \email{jay@eeyore.rice.edu}}
%\altaffiltext{4}{Department of Astronomy, University of Washington, Box
%                 351580, Seattle, WA 98195-1580\\
%                 \email{king@astro.washington.edu}}
%\altaffiltext{5}{INAF-Osservatorio Astronomico di Collurania, via Mentore
%                 Maggini, 64100 Teramo, Italy\\
%                 \email{cassisi@oa-teramo.inaf.it,pietrinferni@oa-teramo.inaf.it}}
%\altaffiltext{6}{INAF-Osservatorio Astronomico di Padova, Vicolo
%                 dell'Osservatorio 5, I-35122 Padua, Italy\\
%                 \email{arenzini@oapd.inaf.it}}

\altaffiltext{1}{Dipartimento di Astronomia, Universit\`a di Padova,
                 Vicolo dell'Osservatorio 3, I-35122 Padua, Italy;
                giampaolo.piotto, sandro.villanova@unipd.it}
\altaffiltext{2}{European Southern Observatory, Karl-Schwarzschild-Strasse 2,
                 D-85748 Garching, Germany; lbedin@eso.org}
\altaffiltext{3}{Department of Physics and Astronomy, Mail Stop 108, Rice
                 University, 6100 Main Street, Houston, TX 77005;
                 jay@eeyore.rice.edu}
\altaffiltext{4}{Department of Astronomy, University of Washington, Box
                 351580, Seattle, WA 98195-1580; king@astro.washington.edu}
\altaffiltext{5}{INAF-Osservatorio Astronomico di Collurania, via Mentore
                 Maggini, 64100 Teramo, Italy;
                 cassisi, pietrinferni@oa-teramo.inaf.it}
\altaffiltext{6}{INAF-Osservatorio Astronomico di Padova, Vicolo
                 dell'Osservatorio 5, I-35122 Padua, Italy;
                 arenzini@oapd.inaf.it}

\begin{abstract}
Accurate photometry with {\sl HST\/}'s ACS shows that the main
sequence of the globular cluster NGC 2808 splits into three separate
branches.  The three MS branches may be associated with complexities
of the cluster's horizontal branch and of its abundance distribution.
We attribute the MS branches to successive rounds of star formation,
with different helium abundances; we discuss possible sources of
helium enrichment.  Some other massive globulars also appear to have
complex populations; we compare them with NGC 2808.
\end{abstract}


%%%
\section{Introduction}
%%%

It had always been considered that the value of studying globular
clusters (GCs) is that they host simple, single stellar populations,
i.e., that stars of a given cluster are not only at the same distance,
but are also coeval and chemically homogeneous.
However, a new set of results has taken the GC population/abundance
field in a new and exciting direction.  Bedin et al.\ (2004, B04)
have found that for a few magnitudes below the
turn-off (TO), the main sequence (MS)
of $\omega$ Centauri splits in two.  The more shocking
discovery, however, came from a follow-up spectroscopic analysis that
showed that the blue MS has twice the metal abundance of the dominant
red branch of the MS (Piotto et al.\ 2005, P05).  The only isochrones
that would fit this combination of color and metallicity were
greatly enriched in helium ($Y\sim 0.38$) relative to the
dominant old-population component, which presumably has primordial
helium.

Indeed, the scenario in $\omega$ Cen is even more complex.  As is
already evident in the color-magnitude diagram (CMD) of B04, this
object has at least three MSs, which spread into a highly multiple
sub-giant branch (SGB) with at least four distinct components 
characterized by different metallicities and ages (Sollima et al.\
2005, Villanova et al.\ 2007; the latter has a detailed discussion.)
These results 
reinforce the suspicion that the multiple MS of
$\omega$ Cen could just be an additional peculiarity of an already
anomalous object, which might not even be a GC, 
but a remnant
of a dwarf galaxy instead (Villanova et al.\ 2007).

In order to shed light on the possible presence of multiple MSs in
Galactic GCs, we undertook an observational campaign with {\sl HST}
this paper we present the first results, on NGC 2808.  D'Antona et
al.\ (2005) had suggested that the MS of this cluster has an anomalous
blueward extension that involves $\sim$ 20\% of the stars.  Here we
will demonstrate that the blueward extension is indeed real, and that,
surprisingly, it is due to the presence of at least three distinct MSs.
We will take advantage of multi-epoch observations to verify, through
proper motions, that these stars are really cluster members, and we
will discuss the implications of this strange population.

%%%
\section{Observations and Measurements}
%%%

For this project we used {\sl HST} WFC/ACS images from GO-9899 and
GO-10922 (P.I.\ Piotto), taken at three epochs:
\\
\phantom{f} May 2005  \hfill     6$\times$340s F475W \phantom{i} \\
\phantom{i} Aug.\ 2006 \hfill 2$\times$350s F475W + 3$\times$350s \
             F814W \phantom{i} \\
\phantom{i} Nov.\ 2006 \hfill 2$\times$360s F475W + 3$\times$360s \
             F814W\phantom{n}\\
The reductions were done with the algorithms described by Anderson \&
King (2006), with one important change:\ whereas they used a
perturbation PSF that is constant across the WFC detector, we found
here that the focus during the 2006 observations was so far off
nominal that a 3$\times$4 array of perturbation PSFs produced better
results.

The photometry was put into the WFC/ACS Vega-mag system according to
the procedure given in Bedin et al.\ (2005), and using the encircled
energy and zero points given by Sirianni et al.\ (2005).  No CTE
corrections have been made, since they would not affect the
% KK: shortened
%photometric results that we discuss here.
present photometric results.
The proper motions were
derived as in Bedin et al.\ (2006).
% KK: not needed in this paper
%, and will be the subject of another paper.

\section{The Color-Magnitude Diagram}
%%%

The left panel of Fig.~1 shows our CMD.  The most surprising result is
clearly visible:\ the MS of NGC 2808 is split into at least three
sequences.  Thanks to our longer color baseline, the split is even
more evident than the one that B04 showed in $\omega$ Centauri.  The
MS complexity is quite different in the two cases, however.  Omega Cen
has, in addition to its principal MS, a less-populated branch on the
blue side and another on the red side, the latter joining to the
anomalous red giant branch, the RGB-a of Pancino et al.\ (2000).  NGC
2808 has neither RGB-a nor a redward branching of its MS; instead it
has {\it two} extra sequences on the blue side of its
MS.  Extending from the dense red part of the MS is only a sprinkling of
stars, in the region where we expect to see MS-MS binaries.

Since NGC 2808 is projected on a rich Galactic field, it is important
to verify that the stars on the additional MSs are really cluster
members.  The middle panel of Fig.~1 shows
the proper-motion distribution, over the 18-month baseline that our
data sets span.  Cluster members furnish the zero point of the
motions.  A few stars have a motion that is clearly different; they
must be the foreground/background objects.  We make a conservative
selection of cluster members with the red circles in Fig.~1.

As shown by Bedin et al.\ (2000, B00), NGC 2808 is affected by a small
amount of differential reddening.  The CMD of the right-hand panel of
Fig.~1 shows the proper-motion-selected cluster members, corrected for
differential reddening with the procedure described in Sarajedini et
al.\ (2007).  A zoomed version of the same CMD is shown in Fig.~2.  An
excess of stars on the blue side of the MS is already evident just
below the TO, while a sequence between the bluest sequence and the
reddest, most populated one starts to be visible at F814W $\sim 20$.
We will refer to these sequences as bMS, mMS, and rMS, respectively,
from left to
right.  It is noteworthy that the three sequences merge close to the
TO, and that for brighter magnitudes the CMD remains narrow, no wider
than expected from the observational errors.  Thus the morphology of
the SGB of NGC 2808
is quite different from that of $\omega$~Cen,
where the multiple SGBs have a vertical extent of more than 1.2
magnitudes (see B04, Sollima et al.\ 2005, and Villanova et al.\
2007).  This difference is important, as it implies for NGC 2808 a
star-formation history
that is significantly different from that of $\omega$
Cen, even though both clusters show multiple MSs.

The splitting of the MS in NGC 2808 is shown quantitatively in Fig.~3.
The left-hand panel shows a part of the right-hand panel of Fig.~1; in
the central panel we have subtracted from the color of the MS the color
of a fiducial line of the mMS, drawn by hand (the continuous line in
Fig.~3).
The right-hand panel shows the color distribution of the points in the
middle panel; the distributions have three clear peaks.

In order to estimate the fraction of stars in each of the MSs, we began
by fitting each of the histograms with three least-squares Gaussians,
% KK: removed words
%We did this fit
independently in each 0.3-magnitude interval from F814W
% KK: removed word
%magnitude
19.5 to 22.5.  We then iterated as follows.  We drew the lines
that are shown; the dashed line runs $3 \sigma$ on the red side of the
rMS, while the dotted line farther to the right is the locus of
equal-mass binaries made of rMS stars.  We then did a new fit of the
Gaussians, but this time we excluded from the fit all stars to the red
of the dashed line, because all those stars are likely to be binaries,
or else field stars that accidentally are close to the cluster motion.
(The continuous red lines show the fits.)  Adding up the areas under the
10 Gaussians for each sequence, we found that $13\pm 5$\% of the stars
belong to the bMS, $15\pm 5$\% to the mMS, and $63\pm 5$\% to the rMS.
The remaining 9\% of the stars are binaries or unremoved field stars.

% KK: shortenened text
%We have mentioned that NGC 2808 is affected by a small amount of
%differential reddening, and that we have corrected our CMD for it.  We
%cannot remove residual reddening variations at very small spatial
%scales, but
We have corrected our CMD as well as we can for differential reddening.
There are at least three reasons for excluding the possibility that the
multiple MSs of NGC 2808 could be a consequence of remaining spatial
variations of the reddening: 1) Differential-reddening effects would
also be evident at the level of the TO, where the sequence is almost
perpendicular to the reddening direction, but Fig.~1 shows that in that
region the color distribution of the stars is narrowest.
%GP Added, as Ivan suggested in response to the referee point 1).
(We believe that this is the strongest evidence for the reality of the
MS splitting.)
2) B00 have estimated that the
differential reddening is of the order of $\Delta E(B-V)=0.02$,
corresponding to $\Delta E(m_{\rm F475W}-m_{\rm F814W})\sim 0.036$.  The
MS split at $m_{\rm F814W}=21.5$ is almost an order of magnitude larger
than this.  3) The three distinct sequences can be seen everywhere in the
field.
%
%%%%%%%%%%%%%%%%%%%%%% END

\section{Discussion}

Before discussing the implications of the triple MS, we need to recall
two additional sets of observed facts about this cluster:

1) NGC 2808 has a very complex HB.  First, it is greatly extended; among
Galactic GCs only $\omega$ Cen and M54 have HBs that go so far to the
blue.  Second, the distribution of stars along the HB is multimodal,
with at least three significant gaps (Sosin et al.\ 1997; B00),
one of these gaps being at the color of the RR Lyrae instability strip.
In fact, even though the HB is well populated both to the blue and to
the red of the instability strip, very few RR Lyraes have been
identified in NGC 2808.  The other two gaps are on the blue extension of
the HB, and delimit three distinct segments, which B00 called EBT1, EBT2,
and EBT3.  B00 found that $46\pm10$\% of the HB stars
belong to the red part of the HB, while $35\pm10$\% are in EBT1,
$10\pm5$\% in EBT2, and $9\pm5$\% in EBT3.
D'Antona and Caloi (2004) have suggested that the HB multimodality
could be due to a multimodal distribution of helium abundances.

2) From an analysis of medium-high-resolution spectra of 122
red-giant-branch (RGB) stars, Carretta et al.\ (2006, C06) have found a
significant Na-O anticorrelation in NGC 2808.  The bulk of the stars 
% KK2
%in NGC 2808 
are O-normal,
% KK: removed word
%stars,
with a peak in [O/Fe] at +0.28, but there
are two additional groups,
% KK: removed words
%of stars,
which they call O-poor (peak at
[O/Fe] = $-0.21$) and super-O-poor (peak at [O/Fe] = $-0.73$).  C06 also
found a marginal increase in [Fe/H], from $-1.113\pm0.008$ for 
% KK2: You might dislike this one and want to restore it (2 parts).
%the
O-normal 
%component 
to $-1.079\pm0.014$ for super-O-poor.  They interpret
this as evidence of increased helium; as discussed by B\"ohm-Vitense
(1979), helium enrichment makes the metal lines look stronger.

Thus three pieces of observational evidence (photometry of the HB,
photometry of the MS, and spectroscopy of the RGB) all seem to point in
the same direction:\ NGC 2808 contains at least three different groups
of stars, with little or no dispersion in iron abundance, but with
different ratios of other metals, and different photometric properties.
%GP Changed and added text from here to the end of the paragraph,
%to account for the first part of comment 2 of the referee
%As all stars have basically the same iron content, the only way we see
%to account for the triple MS is to appeal to differences in helium
%abundance.
As all cluster stars have basically the same iron content,
differences in
%GP3
%helium
He abundance seem to be the only way to account for
the triple MS. In the case of the MS splitting in $\omega$~Cen, no
alternative to the
%GP3
%helium
He explanation has been proposed so far, and therefore
% KK2: small stylistic change
%we explore further this possibility.
we explore this possibility further.
%GP3
%for the case of NGC2808.
%GP added to account for point 5 of the referee
Note that the CNO-Na variations, indicated by the presence of the three
[O/Fe] groups, cannot produce the MS split, as isochrones computed adopting
canonical heavy-element proportions or a mixture mimicking an extreme CNO-Na
anticorrelation perfectly overlap on both the MS and the RGB (Salaris et al.
2006). The MS split occurs only if the CNO-Na anticorrelation is accompanied
by an increase of He content.

In order to test this hypothesis,
%GP3
%against our observations,
we computed
stellar models for a metallicity suitable for this cluster, $Z=0.002$,
with $\rm [\alpha/Fe] = +0.4$ and with various values of the He
content
%GP3 (For more details
(see Pietrinferni et al.\ [2006].)  We
adopted a distance modulus $(m-M)_0=15.0$ and
%GP3
%a reddening
an $E(B-V)=0.18$.  In the inset of Fig.~2 we compare our data with
isochrones for age 12.5 Gyr, with $Y=0.248$, 0.30, 0.35, and 0.40.  The
location of the rMS fits the isochrone for the canonical He content of
$Y=0.248$, while the mMS
%GP3
%location
is well matched by the $Y=0.30$
isochrone.  The bMS falls between the $Y=0.35$ and $Y=0.40$ isochrones.
%GP Added to account for the second part of comment 2 and comment 3 of
%   the referee
Note that the 12.5 Gyr isochrones give a good fit of the
turnoff/subgiant part of the CMD, which is quite narrow, leaving no room
for any appreciable age difference between the various
subpopulations. The less satisfactory fit at the faintest magnitudes of
the $Y=0.248$ isochrone is a known problem for
%GP3
%clusters
GCs of this
metallicity (see Bedin et al.\ 2001), and 
% KK2
%it 
is due to the less than
optimal color-temperature transformations for low-mass stars.

NGC 2808 is similar to $\omega$~Cen in having multiple stellar
populations that are distinct, but there are two important differences:
(1) In $\omega$ Cen the He-enhanced MS has an [Fe/H] that is $\sim$0.3
dex larger than that of the main stellar population (P05), whereas in
NGC 2808 C06 find very little difference in [Fe/H] ($\sim$0.03 dex)
between the two extreme RGB populations, O-normal, and super-O-poor.  (2)
The MS of NGC 2808 suggests three different He values, whereas only two
appear to be necessary in $\omega$ Cen (P05, Villanova et al.\ 2007).

What can we say about connections among the various groups of MS, RGB,
and HB stars?  We know that there are three MSs, plus a well-populated
binary sequence, three groups of RGB stars 
characterized by 
different
%GP3
%oxygen
O content, and four distinct HB sections.  Since the rMS
component includes the majority of MS stars, their progeny must also
remain dominant on the RGB and HB, indicating a connection between the
rMS, the O-normal RGB, and the red part of the HB.
Then, since one expects
%GP3
%oxygen
O depletion to go hand in hand with
%GP3
%helium
He enhancement, it is
%GP3
%quite
natural to connect the mMS to the
O-poor RGB, and the bMS to the super-O-poor RGB. Indeed, to get
any significant
%GP3
%oxygen
O depletion, CNO cycling at high temperature must
operate, and the product of CNO cycling is
%GP3
%helium
He.
The connections with the other HB segments are less obvious, however,
as their locations depend on the unknown amount of mass loss that the
stars in each group suffered when they were on the RGB.  In any case,
higher
%GP3
%helium
He implies a bluer HB location and hence tentative
connection of the O-poor and super-O-poor groups to EBT1 and EBT2,
respectively.

But this still leaves us with the fourth HB group, EBT3.  These
high-temperature stars must either have extremely high
%GP3
%helium
He, or else
% KK2
%must have 
very thin envelopes, after extreme mass loss along the RGB.
%GP3
%On the face of it
Therefore, one would be strongly tempted to connect the EBT3
stars to the MS binaries, given that each represents $\sim 9\%$ of the
stars in that part of the CMD.
Binaries can indeed experience enhanced
mass loss on the RGB, e.g.\ due to Roche-lobe mass transfer,
and this might help to populate EBT3.
After all, it has been demonstrated (Maxted et al.\ 2001) that a large
fraction of hot field sdB stars formed in binaries that went through a
mass-transfer phase.  However, only a fraction of MS binaries will
undergo excess RGB mass loss and land at the blue end of the HB;
others will merge to become blue stragglers.  Moreover, other clusters
with a similar population of MS binaries do not exhibit an extremely
hot HB similar to EBT3.  Thus it may be that binaries help to populate
EBT3, but they cannot be the whole story, so basically we remain
without a satisfactory connection of EBT3 to other parts of the CMD.

These tentative connections between the various groups of MS, RGB and
HB stars are summarized in Table 1.  The columns correspond to regions
in the HR diagram of the cluster and the rows to population
components; the connections go across the rows.  The percentages in
each column add to 100\%.
%GP2 Added after the first part of comment 6) of the referee.
% K:
%Errors comes from Poisson statistics.
The errors come from Poisson statistics.
If the connections are correct, the
percentages should be nearly constant across each row.  They are
reasonably so, when allowance is made for Poisson statistics and for
the fact that the HB lifetime should be $\sim$10\% shorter for stars
with the primordial 
% KK2
%helium 
He value than for those with the highest
% KK2
%helium 
He abundance (Renzini 1977).

NGC 2808 and $\omega$~Cen are among the seven Galactic GCs in the
compilation of Pryor \& Meylan (1993) that have masses greater than
$10^6 M_\odot$.  Two other GCs in this heavyweight group, NGC 6388 and
NGC 6441, also show evidence of multiple stellar populations on the HB
(Rich et al.\ 1997), possibly associated with strong
%GP3
%helium
He enhancements (Sweigart \& Catelan 1998, Raimondo et al.\ 2002, Caloi
\& D'Antona 2007).  It is therefore
%GP3
%rather
tempting to speculate that
multiple stellar generations and the super-
%GP3
%helium
He phenomenon are more
likely to appear among the most massive clusters, whose deep potential
wells favor the retention of gas from low-velocity stellar winds.
This fits with the oft-proposed scenario that the
%GP3
%helium
He-rich
populations may have formed out of the
%GP3
%helium
He-rich ejecta from
intermediate-mass AGB stars, $10^8$--$10^9$ yr after the first
%GP3
%(and main)
generation of star birth (see
%GP3
%, e.g.,
Ventura et al.\ 2001).
Note that the presence in NGC 2808 of two groups of stars that are
strongly depleted in O relative to normal GC stars
%GP3
%clearly
indicates a
very efficient CNO process, which would
% KK: shortenened
%give further support to the
further support the
idea of stars formed from material ejected by intermediate-mass AGB
stars.
%GP3 Ivan, can we remove next sentence?
%(We note that of the three remaining massive clusters, one is
%M54, which has a split RGB and has been suggested to be the nucleus of
%a disintegrating dwarf galaxy [Layden \& Sarajedini 2000], but whose
%MS has not yet been examined for splitting.)
% KK2: Yes.  Criterion is: think about what the paper proclaims -- does
% this contribute to that?  No, it is only rambling.

In the uppermost three intervals of Fig.~3, the Gaussians that best fit
the color distributions of the bMS and the mMS have sigmas
of 0.01 mag, consistent with the width that would come from photometric
errors alone.
The sigma of the rMS is twice as large; but since it is contaminated
by binaries, it is hard to say whether there is a real spread.  Thus
there is no clear evidence for a spread in
%GP3
%helium
He content within each
of the sequences.  This makes it more likely that
%GP3
%helium
He accumulation
in the interstellar matter and star formation have been {\it
sequential} events, whereas a spread would have indicated
%GP3
%helium
He accumulation and star formation as {\it concurrent} events.  Even if
this scenario is qualitatively very attractive, it has some serious
quantitative problems.  In some cases a very high
%GP3
%helium
He abundance is
required, and AGB stars could not produce it
(Karakas et al.\ 2006).  Also, the large mass fraction of the
%GP3
%helium\
He-enriched population relative to the enriching one would require
an
%GP3
%extreme, very
extremely flat IMF (see
%GP3
%, e.g.,
D'Antona \& Caloi 2004).
Alternative sources of
%GP3
%helium
He, such as massive rotating stars (Maeder \&
Meynet 2006) can lead to higher
%GP3
%helium
He yields, but their high-velocity
winds would be much harder to retain within a
%GP3
%cluster
GC.
Moreover, such massive stars would explode as
%GP3
%supernovae
SNe, ejecting gas
rather than holding it, and increasing the metallicity beyond what is
observed in NGC 2808.

We conclude that the existence of multiple stellar populations within
some of the most massive GCs is now a well-established fact.
Apparently, the only way we have of interpreting this observational
evidence is to assume that a fraction of GC stars have sizable helium
enhancements over the primordial value.  What remains to be understood,
however, is the sequence of events that could have led to the formation
of such populations, along with the source of the helium enrichment
itself.

\acknowledgements
I.R.K.\ and J.A.\ acknowledge support from STScI grants GO-9899 and GO-10922.
\ \\

%GP3 All references updated (added authors for references with <=8 authors)
% KK2: corrected some small punctuation errors

\clearpage

\begin{deluxetable}{ccc}
\tabletypesize{\scriptsize}
\tablewidth{0pt}
\tablecaption{\scriptsize{The population components of NGC 2808}}
\tablehead{
\colhead{MS} & \colhead{RGB} & \colhead{HB}}
\startdata
\hline
           &             &                  \\
rMS        &  O-normal   &     red segment  \\
63\%$\pm$5 & 61\%$\pm$7  &      46\%$\pm$10 \\
$Y=0.248$  &             &                  \\
           &             &                  \\
mMS        &    O-poor   &      EBT1        \\
15\%$\pm$5 & 22\%$\pm$4  &      35\%$\pm$ 10\\
$Y=0.30$   &             &                  \\
           &             &                  \\
bMS        &super O-poor &      EBT2        \\
13\%$\pm$5 & 17\%$\pm$4  &      10\%$\pm$5  \\
$Y=0.37$   &             &                  \\
           &             &                  \\
binaries   &   ?         &      EBT3?       \\
9\%$\pm$5  &             &      9\%$\pm$5   \\
\enddata
\label{t4}
\end{deluxetable}

\clearpage

\begin{figure}
\plotone{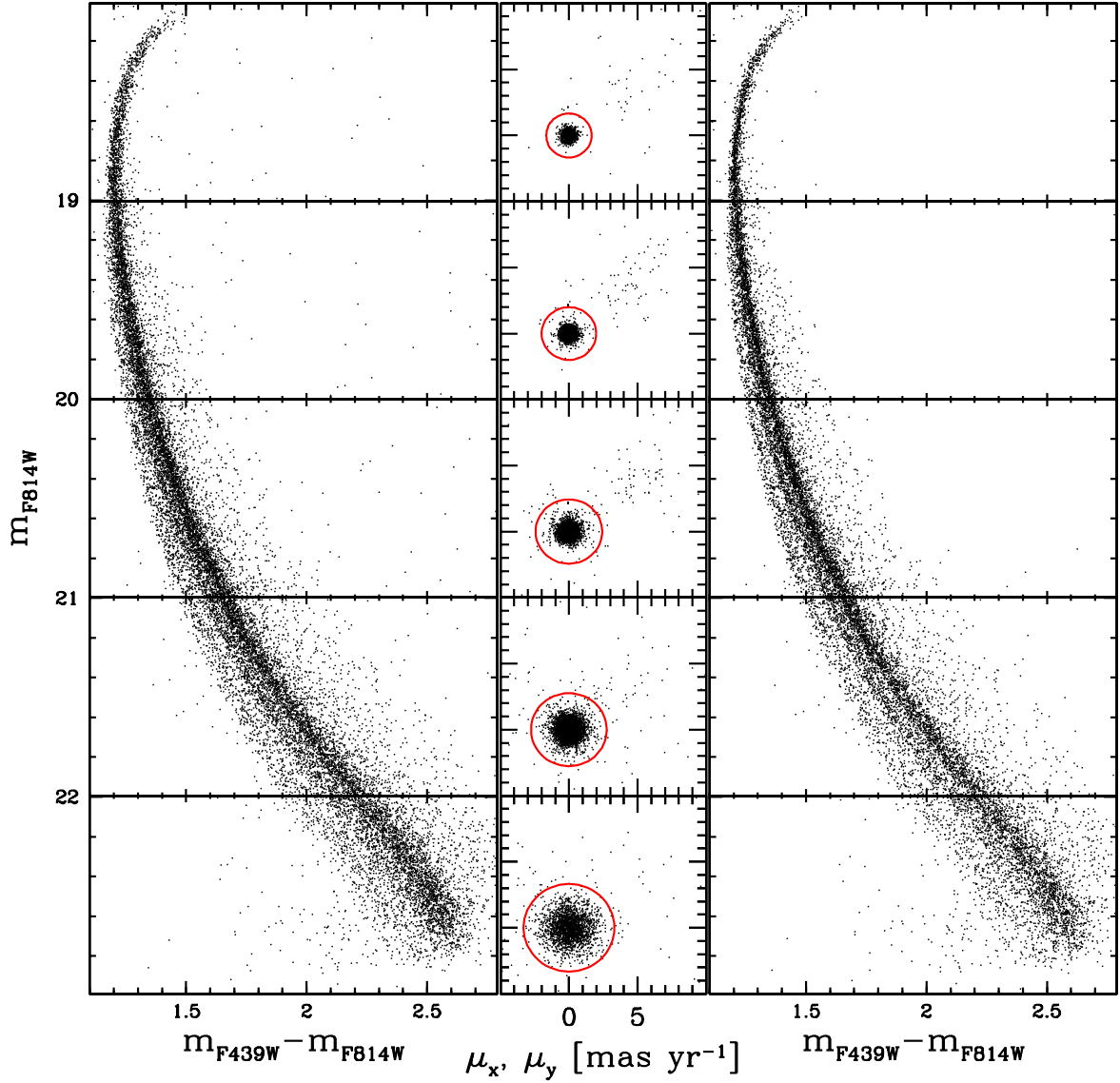}
\caption{Left, the original CMD; middle, the proper-motion distributions
  in the various magnitude intervals;
  right, the CMD of the stars whose
  proper motions lie inside the red circles, with corrections for
  differential reddening.}
\label{f1}
\end{figure}

\clearpage

\begin{figure}
\plotone{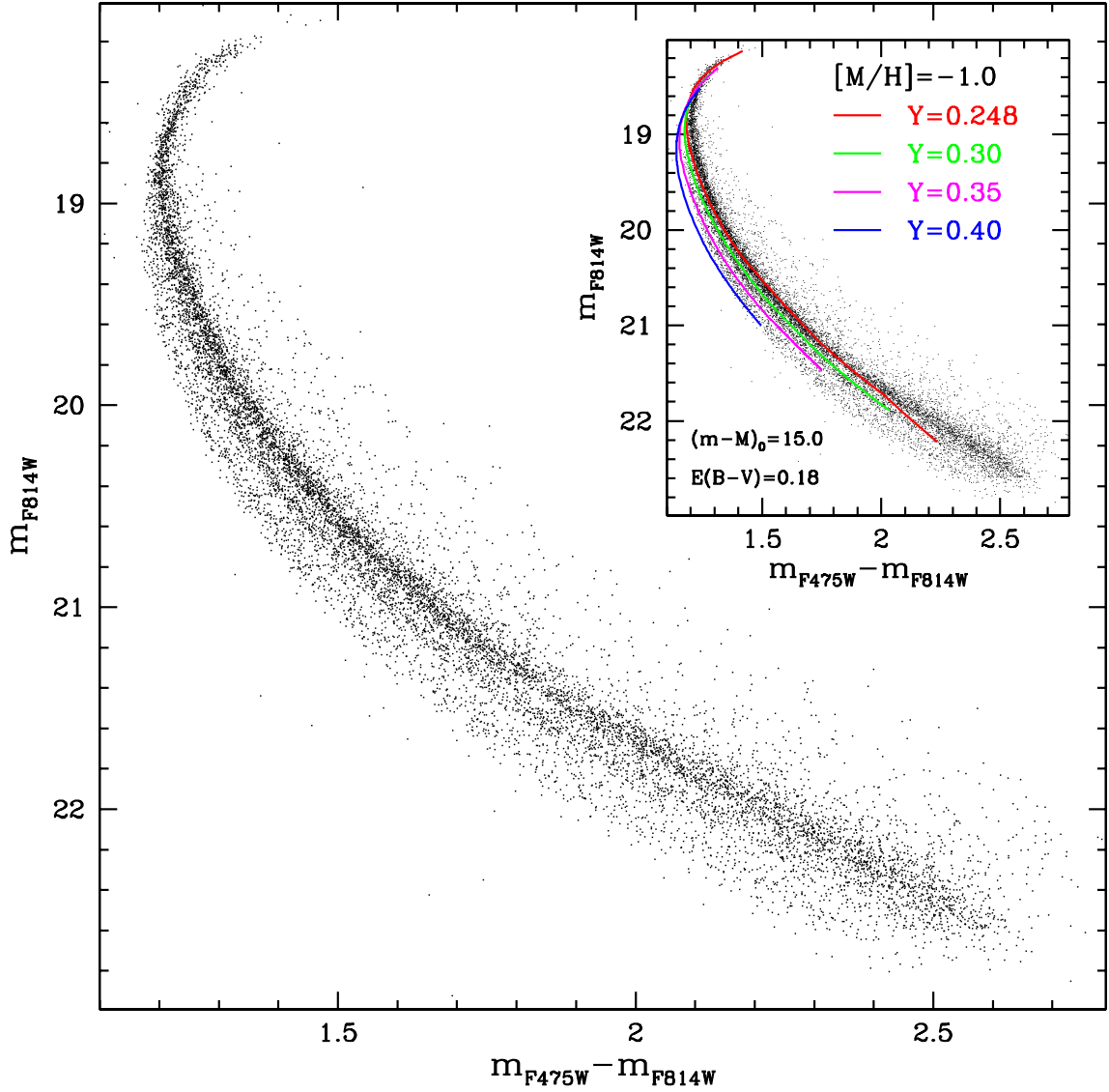}
\caption{A zoom of the proper-motion-selected,
differential-reddening-corrected CMD of the right-hand panel of Fig.~1.
In the inset the observed CMD is fitted with four
%GP2 Added to account for comment 2 of the referee
12.5 Gyr
isochrones, with
different He content.}
\label{f3}
\end{figure}

\clearpage

\begin{figure}
\plotone{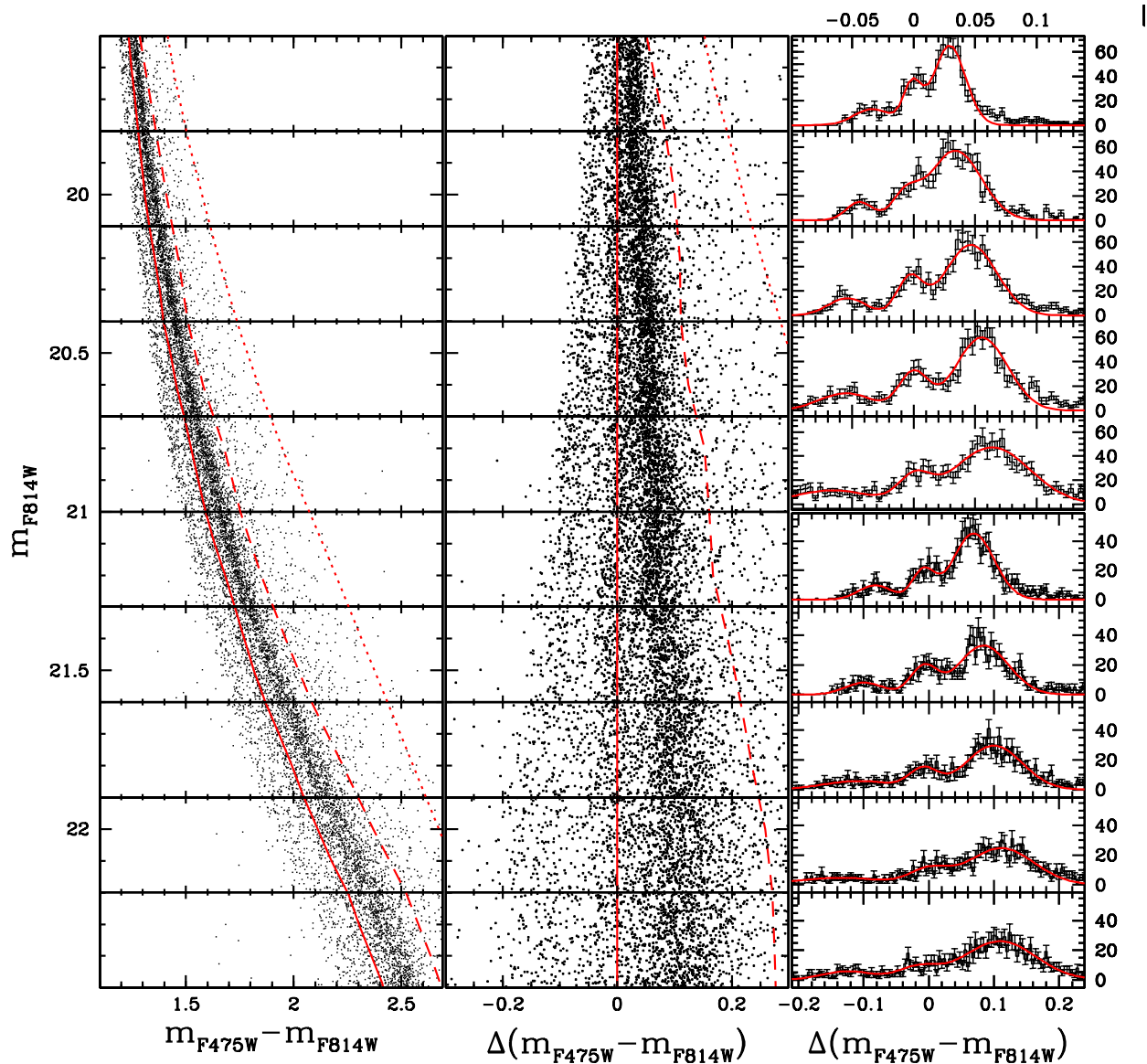}
\caption{Left: The main part of the CMD from Fig.~1.  Continuous line is
a fiducial line of the mMS, drawn by hand.  Dashed line runs $3\sigma$
to the red of the rMS.  Dotted line marks where equal-mass pairs of rMS
stars would lie.  Middle: The same CMD, after subtraction of the color
of the mMS fiducial line.  Right: The color distribution of the points
in the middle panel.
Note that the color scale is different for the upper
half and the lower half.  The continuous red lines are fits by a sum of
three Gaussians.}
\label{f2}
\end{figure}

\end{document}